# Automated Formal Equivalence Verification of Pipelined Nested Loops in Datapath Designs

Payman Behnam, Student Member IEEE, Bijan Alizadeh, Senior Member IEEE, Sajjad Taheri, Student Member IEEE

**Abstract**— The ever-growing complexity of digital systems has made designers move toward using Electronic System Level (ESL) design methodology at a higher abstraction level. The designs at ESL are then automatically synthesized to Register Transfer Level (RTL) by means of High Level or behavioral Synthesis (HLS) tools. Due to possibility of buggy synthesis, especially when the target design must be manipulated or optimized (i.e., pipelining), an efficient equivalence checking method is necessary to check functional equivalency of the ESL specification and the RTL implementation. This problem is even more serious in the case of loop pipelining, since several challenges such as overlapping execution, retiming and forwarding occurre which make traditional sequential equivalence checking approaches, inapplicable. At the same time, the growing market for datapath dominated applications such as DSP for multimedia applications and embedded systems requires a suitable Computer-Aided Design (CAD) support for their verification. In this paper, we present an efficient formal approach to check the equivalence of synthesized RTL against the high-level specification in the presence of pipelining transformations. To increase the scalability of our proposed method, we dynamically divide the designs into several smaller parts called segments by introducing cut-points. Then we employ Modular Horner Expansion Diagram (M-HED) to check whether the specification and implementation are equivalent or not. In an iterative manner, the equivalence checking for each segment is performed. At each step, the equivalent nodes and those nodes which have an impact on them are removed until the whole design is covered. Our proposed method enables us to deal with the equivalence checking problem for behaviorally synthesized designs even in the presence of pipelines for nested loops. The empirical results demonstrate the efficiency and scalability of our proposed method in terms of run-time and memory usage for several large designs synthesized by a commercial behavioral synthesis tool. Average improvements in terms of the memory usage and run time in comparison with SMT- and SAT-based equivalence checking are 16.7× and 111.9×, respectively.

**Index Terms**— Formal verification, equivalence checking, piplined nedted loop, HED

——————————— ◆ ———————————

## 1. INTRODUCTION

The complexity of next generation of digital systems has overtaken traditional time consuming handcrafted RTL design methods. Therefore, some approaches are desirable to generate RTL codes automatically. High Level Synthesis (HLS) tools have been provided to respond to such needs. The HLS is the process of generating RTL design from higher level programs such as C, C++, SystemC, or so on [6]. Using HLS tools leads to more productive designs for next-generation, computationally intensive applications. When we make use of HLS tools, however, we need to make sure that synthesized RTL is bug free. This indicates that the transformation correctness of high level or behavioral synthesis phase is very important [2, 3].

A large amount of work has been done to verify the RTL against its specification. A combinational equivalence checking approach between designs in SystemC and RTL has been suggested in [4]. The authors of [7], [8] and [9] presented Sequential Equivalence Checking (SEC) approaches between software specification and hardware implementation. During equivalence checking, several optimization techniques such as cut-point, cut-plane and cut-loop are used. The cut-point optimization is to find internal equivalent nodes of specification and their corresponding circuit implementations. Cut-points reduce the size of symbolic expressions by replacing verified sub-circuits with new symbolic values [28]. Cut-plane is considered as a set of cut-points while cut-loop is considered as a cut-plane at the end of a loop [30]. Some techniques have been proposed to check the equivalency between combinational circuits with some structural similarities using bit-level decision diagrams [28, 31] SAT-based approaches [38, 39, 43, 44], probabilistic methods [46] and directed test generations [47]. The structural similarities enable them to find identical internal nets as cut-points to partition the whole design into a set of smaller segments. However, their approach to problem of equivalence checking is limited to bit level verification and hence cannot handle large RTL designs. In [30], the authors have proposed a novel approach to verify equivalence of C-based system level description versus RTL model by looking for merge-points as early as possible to reduce the size of equivalence checking problems. This method however is suggested for hand crafted RTL codes. In addition, it makes use of cut-loop techniques which are inapplicable to pipelined designs as will be discussed in Section 2.

Lots of work has been performed for equivalence checking of generated RTL using HLS tools against its specification [5, 10, 11, 12, 13, 14]. The authors of [5] have

———————————————

*Payman Behnam is with the Computer Science Department, University of Utah, SLC, USA. Bijan Alizadeh is with the Electrical and Computer Engineering Department, University of Tehran, Tehran, Iran. Sajad Taheri is with the Computer Science Department, University of California Irvine, Irvine, USA (e-mail: payman.behnam@utah.edu, b.alizadeh@ut.ac.ir, sajjadt@uci.edu).*

used a bi-simulation correspondence checking to validate designs generated by the SPARK behavioral synthesis tool. A suite of optimizations for the SEC framework has been presented by [10] which exploit both the explicit control and data flow representations in the Clocked Control and Data Flow Graph (CCDFG) and the module structures in the ESL description. The authors of [12] proposed a SEC framework to compare an ESL design with its behaviorally synthesized RTL in the presence of optimizations such as operation gating and global design variables. The work in [13] has tried to solve the equivalence checking problem for compiler transformations in behavioral synthesis.

The process of behavioral synthesis consists of several transformation phases including compilation, scheduling, allocation, binding, and control generation [6, 18]. On the other hand, when the target design must be pipelined, loop pipelining is employed as part of scheduling and binding phases which results in several challenges, such as overlapping execution, retiming, forwarding, and losing direct one to one mapping between the specification and the pipelined RTL. Hence, traditional sequential equivalence checking approaches are becoming inefficient. Despite the existence of numerous methods to verify pipelined microprocessors [15, 16, 17], there are a few published approaches on formal equivalence checking of behaviorally synthesized pipelined loop designs.

The comparison of input-output relations between the specification and the high level synthesized pipelined RTL is prohibitively expensive for loops with many iterations. A reference pipelining transformation on the CCDFG was proposed in [11] to deal with the problem of loop pipelining without using approaches based on input-output comparison. The proposed method is based on building reference pipeline model with a certified specific transformation and checking the equivalence between the reference model and synthesized RTL using dual-rail symbolic simulation. However, it requires several parameters whose values are needed to be obtained from the HLS tool. Additionally, although it can handle SEC of pipelined designs with nested loops, it cannot make use of proposed SEC optimization techniques for internal loops and hence has to unroll internal loops. The authors of [14] solved the problem of equivalence checking for function pipelining (instead of loop pipelining) in behavioral synthesis.

Tackling the equivalence checking problem of high level synthesized designs, requires a scalable representation model. In recent years, a strong and scalable high-level decision diagram called M-HED has been proposed [25, 27, 33]. This decision diagram has a compact and a canonical form, and is also close to high-level descriptions of a design. The other properties of M-HED such as a facility for expressing primary outputs of a design in terms of primary inputs in a polynomial form, presenting state variables in terms of integer equations in a formal model, and availability of arithmetic operations in a word-level, has made it a powerful and scalable platform for verification [25, 27, 33, 37, 41, 42].

In this technical report, we present a scalable formal equivalence checking methodology for pipelined loop designs synthesized at high level while nested loops are pipelined and no information from the HLS tool is necessary. The short version of the presented method has been published in [45] as a A ASP-DAC paper. In this version, we describe our proposed method in more details with several examples. As Figure 1 shows, first we perform a symbolic simulation to create a list of assignments of specification (LAASC) and behavioral synthesized design when all loops are pipelined (LAPRTL). These lists completely describe the behavior of specification and implementation. For equivalence checking, we employ an efficient canonical hybrid bit- and word-level decision diagram called M-HED [25, 27, 33]. The key idea to increase the scalability of our framework is to avoid cost-prohibitive input-output comparison by introducing dynamic cut-points instead of fixed cut-points used in [28].

Although the authors of [28] used BDDs and cut-loops for equivalence checking, their method is not able to handle large designs and those designs that contain pipelined loops. As we stated before, loop pipelining is one of the most intricate transformations [40] that arises challenges in developing automated equivalence checking methods. In addition, unlike the method presented in [11], these cut-points are not operation mapping between the specification and the RTL design. Moreover, in contrast with [11], our proposed method doesn't need several parameters whose values are obtained from HLS tool and also doesn't make use of sophisticated concepts of formal verification such as theorem proving for certification framework. This property allows other researchers to understand and replicate it easily. The cut-points in our method are inserted into parts of the code where the size of its corresponding M-HED is maximized. This way, the design is divided into several smaller segments. Therefore, the size of the equivalence checking problem significantly reduces. We continuously check the equivalence of corresponding segments from specification list and pipelined loop synthesized implementation list to detect the equal nodes. Then, we cut out the equivalent nodes and introduce them as new primary inputs for the rest of the segments. These primary inputs are used while next segments should be checked. If no match is found while comparing segments, we make use of an internal equivalence approach that enables us to incrementally remove outputs of corresponding segments, expose temporary nodes as new output nodes, and explore them to check whether they are equivalent or not.

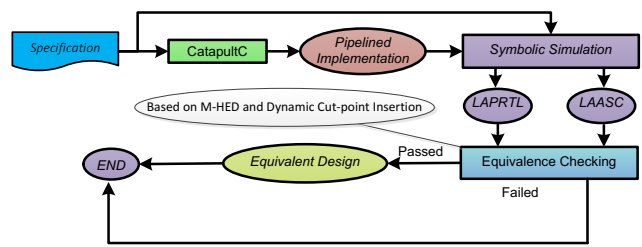

Figure 1. Proposed equivalence checking methodology.

We employ CatapultC [1] as an HLS tool to automati-cally generate thousands of lines of RTL code from C++ code description of several large-sized designs. Although the individual parts of our presented technique are known, our innovation is to employ the unique combination of them. Hence, the main contributions of this paper are as follows:

- Creating a framework based on M-HED for fully automating SEC of complex design where *pipelined nested loops* are supported. To the best of our knowledge it is the first work which utilizes high level decision diagrams for equivalence checking of high level synthesized pipelined loop in data path designs.
- Proposing a method to *overcome the lack of optimization methods* for equivalence checking of pipelined loop designs synthesized at high level. In the proposed method, we can handle the problem of equivalence checking in a reasonable run time in the complex data path designs with pipelined nested loops.
- *Dynamically* specifying the *cut-points* so that pipelined loops can efficiently be verified.

The rest of this paper is structured as follows. The challenge of loop pipelining in equivalence checking is discussed in Section 2. The M-HED as a hybrid canonical representation is expressed in Section 3. Our proposed equivalence checking approach with a simple example is described in Section 4. Experimental results are reported in Section 5, while a brief conclusion is presented in Section 6.

## 2. LOOP PIPELINING CHALLENGES IN EQUIVALENCE CHECKING

In this section, we explain the critical challenges of loop pipelining in verification and equivalence checking when a code in high level language such as C is synthesized into RTL. Loop pipelining is an operation that allows the next iteration of a loop is started before the previous one is fully finished. This operation increases parallelism and throughput. In HLS tools, loop pipelining can be controlled by the number of cycles that must elapse between two successive iterations. This parameter is called initiation interval [1]. For example, consider a pair of three-level nested loops shown in Figure 2. The result of a single iteration of the outer loop when none of the middle and inner loops is pipelined is shown in Figure 3. When initiation interval is one, the result of a single iteration of outer loop when both inner and middle loops are pipelined is shown in Figure 4(a). Finally, Figure 4(b) illustrates the result of single iteration when all loops are pipelined while initiation interval is one.

As shown in Figures 3 and 4, when loops are pipelined, the overlapping execution of consecutive iterations appears. It means that i+1$_{th}$ iteration can be initiated before (or concurrent with) i$_{th}$ iteration is committed. These retiming and out-of-order executions cause the sequence of operations in the specification and generated RTL as well as controlling finite-state machines to become different. Hence directly using of several optimizations such as cut-loop techniques becomes inapplicable that makes the problem of equivalence checking difficult.

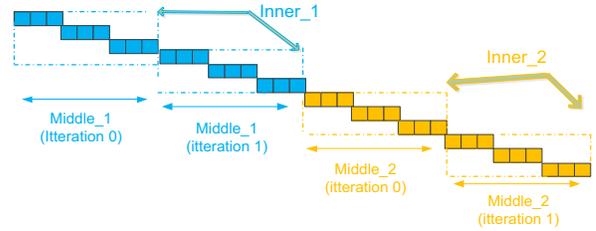

*Outer: For (...) {*
  *middle_1: For ( int i = 0; i < 2; i++)*
    *inner_1: For ( int j = 0; j < 3; j++)*
      *a1[j] = b[i]-c[i]+d[i]-e[i];*
  *middle_2: For ( int k = 0; k < 2; k++)*
    *inner_2: For ( int l = 0; l < 3; l++)*
      *a2[l] = b[k]-c[k]+d[k]-e[k];}*

Figure 2. An example of a pair of three-level nested loop.

Figure 3. Execution order of three-level nested loops with no pipeline

To give a glimpse about the kind of challenge we face with when a C code is synthesized into RTL code, let us consider the C code and related RTL model illustrated in Figure 5. The RTL schematic is obtained by CatapultC[1] when the specification is synthesized at the frequency of 100 MHz. In each iteration, four multiplications, two additions and lastly one multiplication must be performed. In this case, one can use fixed cut-loop optimization technique for equivalence checking. That is because an iteration of specification equals to two cycles of implementation.

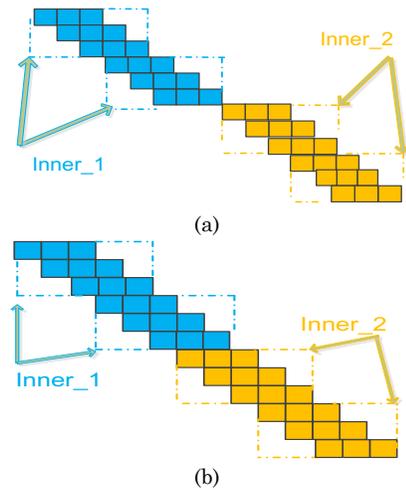

Figure 4. Execution order of tree-Level nested loops (a) when middle and inner loops are pipelined, (b) when outer, middle, and inner loops are pipelined.

In other words, the symbolic values on a cut in the specification always equal to the symbolic values of a specific cut in the implementation. However, when a design is pipelined such a correspondence is lost and it would not be possible to determine fixed cut-loops. In the fixed cut-loop we must be able to find a location in such a way that always after a fixed number of cycles in the implementation, all points in the specific cut of implementation and specification become equivalent.

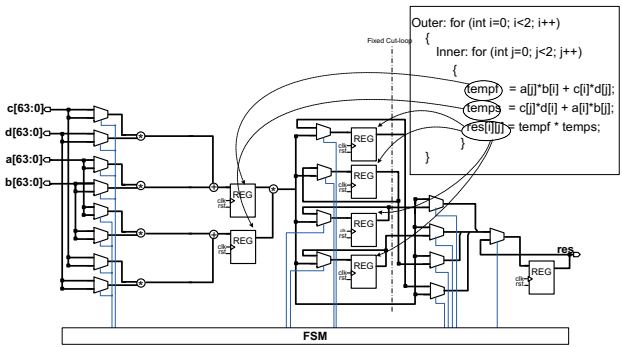

Figure 5. Using a high level synthesis tool (CatapultC) to generate RTL without pipelining nested loop.

Suppose that five multipliers and two adders are available during the high level synthesis phases. Hence, four multiplications in the second iteration of the inner loop (Figure 5) are being computed before the computation of the first iteration is finished. Such reordering deprives us of the opportunity to select a fixed cut-loop in the implementation (cutIi; i=1 to 6 in Figure 6) and the specification (cutSi; i=1 to 6 in Figure 6) in such a way that a regular behavior in the ordering of operations in the implementation and the specification is observed. For example, suppose that we want to find a corresponding cycle in the implementation for the first iteration of the specification (i.e., cutS3) and check whether this cycle can be used as a fixed cut-loop or not. By moving forward in successive cycles of the implementation, cutI3 is found. But several multiplication and addition operations in cutI3 have no corresponding operations in the cutS3 of specification. Based on definition of cut-loop in Section 2, the cutI3 cannot be considered as a cut-loop because no equivalent statement for other statements in cutI3 can be found in cutS3. With proceeding in remaining cycles of the implementation we observe that neither cutI3 nor any other cuts can be used as a cut-loop in a way that always after fixed number of cycles all statements in that cut become equivalent to all statements in a specific cut in the specification. This example shows that in the case of pipelined loop implementation, using optimization techniques such as cut-point, cut-loop, or cut-plane for equivalence checking purposes are not straightforward. These challenges motivate us to come up with an efficient methodology for equivalence checking between the specification and synthesized pipelined loop implementation.

## 3. MODULAR HORNER EXPANSION DIAGRAM (M-HED)

In order to make this paper self-contained, we introduce a graph-based representation called Horner Expansion Diagram (HED) for functions with a mixed Boolean and integer domain, and an integer range to represent arithmetic operations at a high level of abstraction [25, 33]. By contrast, other Word Level Decision Diagrams (WLDDs) are graph-based representations that provide a concise representation of integer-valued functions defined over binary variables as a bit vector. On the other hand, Binary Decision Diagrams (BDDs) or Satisfiability (SAT) based methods suffer from size explosion problems when designs grow in size and complexity. BDD-based verification tools have not been very successful for designs containing large arithmetic data-path units due to prohibitive memory requirements. In HED, we assume that the set of variables is totally ordered and all vertices that have been constructed obey this ordering.

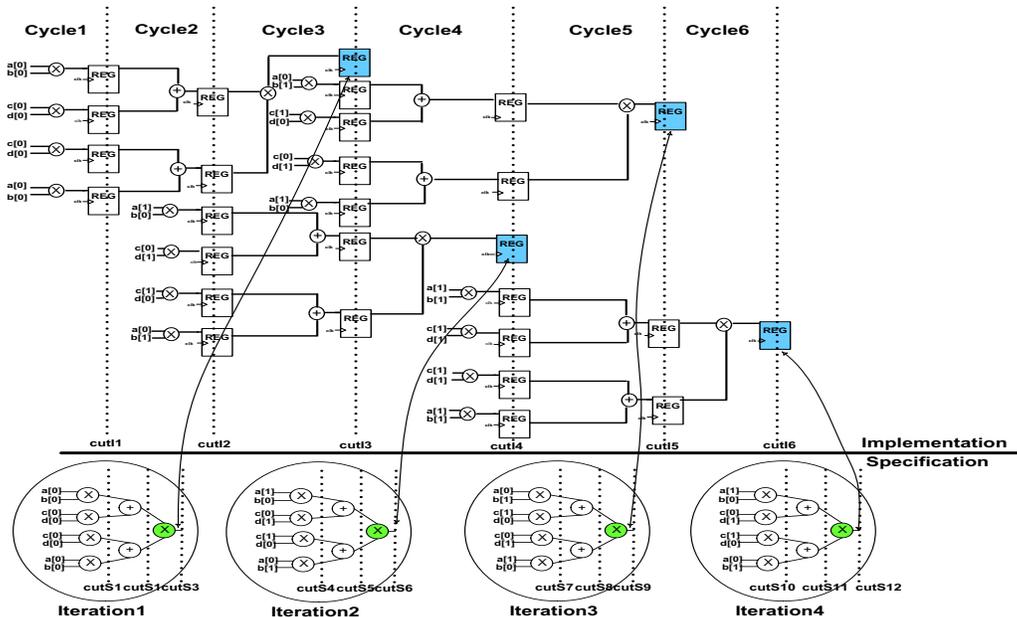

Figure 6. Demonstration of inability to use fixed cut-loops in order to check the equivalence of specification and pipelined RTL implementation when loops are unrolled.

These conventions are similar to other word-level canonical representations and are not discussed here for brevity [21,23]. The HED is a binary graph-based representation where the algebraic expression F(X,Y, …) is expressed by a first-order linearization of the Taylor series expansion [30]. Suppose variable X is the top variable of F(X,Y, …). Equation (1) shows F(X,Y, …), where *const* is independent of the variable X, while *linear* is the coefficient of variable X.

$$F(X, Y, …) = F(X=0, …) + X \times [F'(X=0, …) + …] = const + X \times linear \quad (1)$$

The HED is a directed acyclic graph G = (VR, ED) with vertex set VR and edge set ED. While the vertex set VR consists of two types of vertices: Constant (C) and Variable (V), the edge set indicates integer values as weight attribute. A Constant node v has as its attribute a value $val(v) \in Z$. A Variable node v has as attributes an integer variable $var(v)$ and two children $const(v)$ and $linear(v) \in \{V, C\}$.

### 3.1 Reduction Rules and Canonicity

Analogous to BDDs and *BMDs, HED can be reduced by removing redundant nodes and merging isomorphic nodes. In order to do so, the following reduction rules have been employed:

**Rule 1:** Remove a node if its *linear portion* (right child) is *0-terminal* or its right edge has 0 weight. Then replace this node with its *const portion* (left child). Figure 7(a) illustrates this situation where node *v* contains only const part and therefore the function computed at that node is independent of variable *var(v)*.

**Rule 2:** Merge isomorphic nodes. This merging rule identifies isomorphic sub-graphs. Two nodes are isomorphic if they not only have the same *const* and *linear* portions but also their variable IDs should be the same. Figure 7(b) shows that nodes *v1* and *v2* are isomorphic and then merged together as a new node *v*.

Figure 8 illustrates how f(x, y, z) = 24-8z+12y+12yz-6x-6$x^2$z is represented by the *HED*. Let the ordering of variables be x > y > z. First the decomposition w.r.t. x is taken into account. As shown in Figure 8(a), after rewriting f(x, y, z) = (24-8z+12y+12yz) + x(-6-6xz) based on (1), *const* and *linear* parts will be 24-8z+12y+12yz and -6-6xz, respectively. The *linear* part is decomposed w.r.t. variable *x* again due to $x^2$ sub-monomial. After that, the decomposition is performed w.r.t. variable *y* and then *z* as shown in Figure 8(b). In order to reduce the size of the *HED* representation, redundant nodes are removed and isomorphic sub-graphs are merged. In Figure 8(b), 24-8z, 12+12z and -6z are rewritten by 8[3+z(-1)], 12[1+z(1)] and -6[0+z(1)], respectively. In order to normalize the weights, gcd(12,12) = 12, gcd(8,12) = 4 and gcd(-6,-6) = -6 are taken to extract common factors. Finally, Figure 8(c) shows the normalized graph where gcd(4,-6) = 2 is taken to extract the common factor between out-going edges from *x* node. In this representation, dashed and solid lines indicate *const* and *linear* parts, respectively. Note that in order to have a simpler graph; paths to 0-terminal have not been drawn in Figure 8(c).

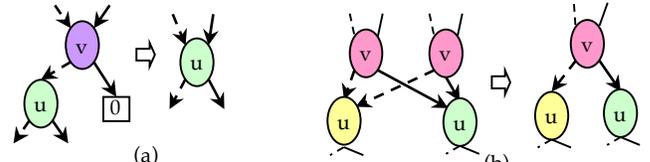

Figure 7. Reduction rules: (a) redundant nodes, (b) isomorphic sub-graphs.

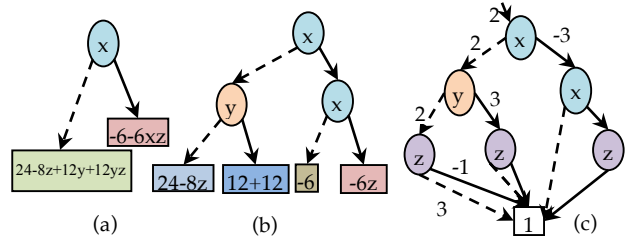

Figure 8. HED representation of 24-8z+12y+12yz-6x-6$x^2$z: (a) decomposition with respect to variable x, (b) decomposition with respect to variables x and y, (c) decomposition with respect to variables x, y and z.

In this graph basic arithmetic operators such as addition, unary addition, subtraction, unary subtraction and multiplication are available that work for symbolic integer variables. In order to represent Boolean functions, logical bitwise operations including NOT, AND, and OR have been provided that are discussed in the following subsections [32, 33].

### 3.2 Boolean Logic

In order to have an integrated representation of Boolean and integer variables, the logical operations need to be supported as well as arithmetic expressions. Boolean operations are defined based on the arithmetic operations illustrated in Figure 9. To address bit-slicing problem, in [32] we have introduced a hybrid method to check the equivalence between a high level specification and the RTL implementation. In our hybrid equivalence checking approach, we have proposed a decomposition technique as shown in Figure 10 that enables us to deal with bit-slicing problem. The main idea is those word-level variables whose bit-slices are used in the Boolean part of the design are decomposed into other word-level variables. For example, if $i^{th}$ bit of variable *Tmp*, i.e. *Tmp[i]*, is used in the Boolean part, the related word-level variable, i.e., *Tmp*, is decomposed into other word-level variables according to Tmp = $2^{(i+1)} \times Tmp_H + 2^i \times y + Tmp_L$, where $Tmp_H$ and $Tmp_L$ are new integer variables and *y* is *Tmp[i]*.

### 3.3 Arithmetic Operations

Addition (W = X+Y) and subtraction (W = X-Y) can be represented canonically based on decompositions in (2) and (3) respectively, as shown in Figure 11(a) and Figure 11(b):

$$W = X + Y = Y + X * (1) = [0 + Y * (1)] + X * (1) \quad (2)$$

$$W = X - Y = -Y + X * (1) = [0 + Y * (-1)] + X * (1) \quad (3)$$

Multiplier (W = X*Y) can be represented canonically based on (4), as shown in Figure 11(c).

$$W = X * Y = 0 + X * (Y) = 0 + X[0 + Y * (1)] \quad (4)$$

```
1) NOT X ==> 1-X
2) X AND Y ==> X*Y
3) X OR Y ==> X+Y-X*Y
```

Figure 9. Logical operations in HED graph.

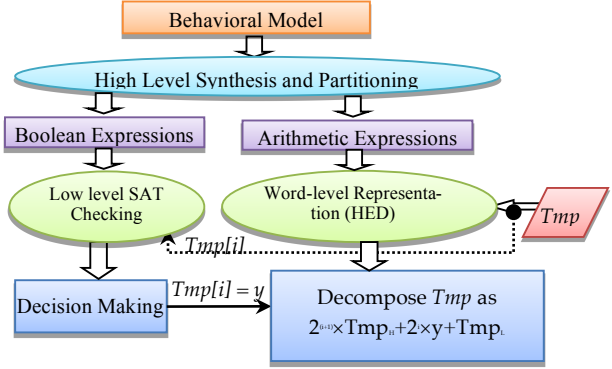

Figure 10. Decomposition technique to address bit-slicing problem.

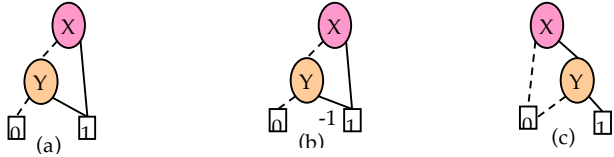

Figure 11. Arithmetic operations in HED: (a) addition, (b) subtraction, (c) multiplication.

In order to describe how arithmetic operators such as addition and multiplication are applied to two *HED* nodes and as a result a new node is generated, let $u$ and $v$ be two nodes to be composed, resulting in a new node $q$. Let $var(u) = x$ and $var(v) = y$ denote the decomposing variables corresponding to the two nodes to be decomposed. The following cases should be considered:

1- If both nodes are *Constant* nodes ($u, v \in C$), a new *Constant* node $q$ is computed as follows:
Addition $q = u + v$: $val(q) = val(u) + val(v)$
Multiplication $q = u * v$: $val(q) = val(u) * val(v)$

2- If one of the nodes is *Constant* node ($v \in C$), a new *Variable* node $q$ is created as follows. For addition operation, the *const* part of result, i.e. $q_0$, is obtained by adding constant $val(v)$ to *const* part of u ($u_0$). The *linear* part of result will be *linear* part of u ($u_1$). For multiplication operation, $val(v)$ should be multiplied with both *const* and *linear* parts of u ($u_0$ and $u_1$) to create *const* and *linear* parts of the result ($q_0$ and $q_1$) respectively.
Addition $q = u + v$: $q_0 + x * q_1 = (u_0 + val(v)) + x * u_1$
Multiplication $q = u * v$: $q_0 + x * q_1$
$= u_0 * val(v) + x * u_1 * val(v)$

3- If both nodes are *Variable* nodes ($u, v \in V$), proceed according to variable order. Suppose $order(x) > order(y)$.

a. Where the two nodes are indexed by different variables, $var(q) = max(var(u), var(v)) = x$. For addition operation, the *const* part of result, i.e. $q_0$, is computed by adding $v$ node to *const* part of u ($u_0$). The *linear* part of result will be *linear* part of u ($u_1$). For multiplication operation, $v$ node should be multiplied with both *const* and *linear* parts of u ($u_0$ and $u_1$) to generate *const* and *linear* parts of the result ($q_0$ and $q_1$) respectively.
Addition $q = u + v$: $q_0 + x * q_1 = (u_0 + v) + x * u_1$
Multiplication $q = u * v$: $q_0 + x * q_1 = u_0 * v + x * u_1 * v$

b. Where the nodes have the same index then $var(q) = x$. In this case, the *const* part of $q$ is created by pairing the *const* parts of two nodes ($u_0 + v_0$ for addition and $u_0 * v_0$ for multiplication). The *linear* part of $q$ is obtained as a sum of two cross products of *const* and *linear* parts when multiplication should be done. Furthermore, the quadratic term, i.e. $q_2$, is taken into account in linear portion of $q$.
Addition $q = u + v$: $q_0 + x * q_1 = (u_0 + v_0) + x * (u_1 + v_1)$
Multiplication $q = u * v$: $q_0 + x * (q_1 + x * q_2)$
$= u_0 * v_0 + x * (u_1 * v_0 + u_0 * v_1 + x * (u_1 * v_1))$

### 3.4 Shift Operations

While shift left operator, <<, can be viewed as scalar multiplication, shift right operator, >>, can be modeled as a division by 2N. In order to compute the division on HED, while assuming the divisor is a constant integer number and also powers of two, 2N, the following recursive algorithm is applied until terminal cases are reached. At terminal nodes, the division is converted to a numerical division problem which is performed easily. If, however, constant values at terminal nodes are less than 2N, the related variable, Var, is replaced by Var/2N [34].

$Z = f / 2^N = (f_{const} + X * f_{linear}) / 2^N = (f_{const}/2^N) + X * (f_{linear}/2^N)$
If terminal $< 2^N$: replace Var by Var/$2^N$

### 3.5 Conditional Statements

In order to handle conditional statements such as if-then-else statement and case statement, variables from different branches of conditional statements are rewritten by different indices (e.g., variable n is defined as n1, n2, … , nm variables for m cases to consider both if and else parts in different iterations). Then these new variables are added to the design in place of conditional statements as shown in Figure 12.

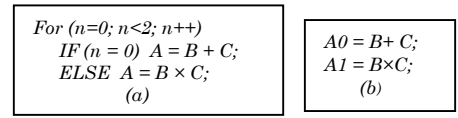

Figure 12. Conditional statements (a) original source code (b) list of assignments

### 3.6 Modular Horner Expansion Diagram (M-HED)

In order to verify polynomial data paths over bit-vectors, we have extended the HED to manipulate modular arithmetic [25]. Although the equivalence verification over $Z_m$ is known to be NP-hard when m≥2 [29], analyzing polynomials over arbitrary finite integer rings and their properties are useful to deal with the equivalence checking problem [20]. The theory of univariate vanishing polynomials over $Z_m$, $m \in N$, $m>1$; i.e., those polynomials $f$ such that $f(x) \mod m \equiv 0$, has been presented in [24]. The authors of [26] have extended the concepts of the work [24] and derived a unique form representation of a multivariate polynomial over finite integer rings of the form

$Z_{p^n}$, where $p$ is any prime integer. Let us consider a simple example that defines functions $f_1[3:0] = 15(Y[3:0])^3 - 5(Y[3:0])^2 + 19Y[3:0] + 6$ and $f_2[3:0] = 7(Y[3:0])^3 + 3(Y[3:0])^2 + 3Y[3:0] + 6$. While $f_1$ is not equivalent to $f_2$ as polynomial functions over Z, they are equivalent over $Z_{2^4}$, i.e. $f_1 \bmod 2^4 \equiv f_2 \bmod 2^4$. Computing their difference over $Z_{2^4}$ results in $f_1[3:0] - f_2[3:0] = 8(Y[3:0])^3 - 8(Y[3:0])^2 + 16Y[3:0]$. While the result is non-zero polynomial, $(8Y^3 - 8Y^2 + 16Y) \bmod 16 = 0, \forall Y \in \{0,1,...,15\}$ and we say $8Y^3-8Y^2+16Y$ vanishes in $Z_{2^4}$. In general, it is not straightforward at all to see whether given polynomials are vanishing ones or not. Actually, a straightforward approach which expands everything into Boolean domain does not clearly work.

We have discussed how the properties of polynomial functions over finite integer ring allow us to reduce two polynomial functions to a canonical form based on the HED [25]. We follow the basic idea proposed in [22] but make use of the HED for efficient implementations and manipulations of polynomials with fixed bit-width that results in a new package called M-HED. Therefore, equivalent polynomial functions over finite integer rings, i.e., data paths with finite bit-width in hardware designs, are automatically identified due to the canonical representation of the M-HED. Since we only use the M-HED to check the equivalence between two polynomials over finite word-length, and manipulating polynomials with fixed bit-width is not our contribution in this paper, we will not discuss further details and we refer the interested reader to our previous work [25].

## 4. PROPOSED EQUIVALENCE CHECKING APPROACH

In this section we explain our methodology in more details. First, it is worthwhile explaining how a specification and an RTL implementation with pipelined loop are converted to a list of assignments, so that they can be represented by M-HED.

### 4.1 Symbolic Simulation

Algorithmic Specification in C *(ASC)* and RTL with pipelined loop in Verilog generated by CatapultC *(PRTL)* are treated as inputs of *SEC-PIPED* algorithm. Then execution of specification and implementation are translated into several assignments by using symbolic simulation (*SymSim* function in lines 1-2 of Figure 15). In symbolic simulation, symbolic values rather than concrete ones (integer or binary values) are used as input vectors. As a clarifying example, consider the C code of 4-point FFT shown in Figure 13. The intent of this code sequence is to perform the butterfly computations with three main loops. The outside loop counts through 2 stages of the FFT computation and it causes huge data-dependent computations. The inner loops perform the individual butterfly computations of each stage. The heart of the FFT algorithm is the block of the code that performs each butterfly computation in the third loop. Note that *wi* ad *wr* parameters are commonly known as twiddle factors and can be computed before the algorithm is fulfilled. To symbolically simulate such a code, the loops are unrolled and therefore a list of assignments is obtained. Then, controlling assignments are removed and the indexes of the arrays are adjusted. Figure 14 illustrates the list of assignments after removing controlling assignments and adjusting the indexes of the arrays in such a way that multiple assignments to a single variable don't happen while data dependencies are preserved. The result of symbolic simulation is a list of assignments which exactly mimics the behavior of given C code. In a similar manner, RTL codes are symbolically simulated.

Note that in the list of assignments, we have three variable types; inputs which are appearing only on the right hand side, outputs which are only in the left hand side and intermediate signals which are in both sides. The objective of equivalence checking is to check whether the functionality of the output variables in the generated assignment list according to the specification and implementation are equivalent or not.

```
//Arrayswr(i)=cos(i*2*PI/N)*512&wi(i)=sin(i*2*PI/N)*512
len = 4;
incr = 1;
for (stage = 0; stage < 2; stage++)
    len1 = len;
    len = len/2;
    windex = 0;
for (j=0; j < len; j++)
    C = wr[windex];
    S = wi[windex];
//butterfly computation
for (index = j; index < 4; index = index + len1)
    index2 = index + len;
    tmr = aar[index] - aar[index2];
    tmi = aai[index] - aai[index2];
    aar[index] = aar[index] + aar[index2];
    aai[index] = aai[index] + aai[index2];
    if (windex == 0)
        aar[index2] = tmr;
        aai[index2] = tmi;
    else
        aar[index2] = tmr*C - tmi*S;
        aai[index2] = tmr*S + tmi*C;
        windex = windex + incr;
incr = 2*incr;
```

Figure 13. 4-point FFT specification.

### 4.2 SEC Algorithm using M-HED

As mentioned in Section 2, the main idea is introducing dynamic cut-points (CPs) to deal with reordering and out of order execution problem. In addition, CPs divide the list of assignments into several smaller segments making large designs tractable by M-HED.

| | |
|---|---|
| len = 4; incr = 1; | aai[3] = tmr*S + tmi*C; |
| //stage = 0 | windex = 2; incr = 2; |
| len1 = 4; len = 2; windex = 0; | //stage = 1 |
| //j=0 | len1 = 2; len = 1; windex = 0; |
| C = wr[0]; S = wi[0]; | //j=0 |
| index = 0; index2 = 2; | C = wr[0]; S = wi[0]; |
| tmr = aar[0] - aar[2]; | index = 0; index2 = 1; |
| tmi = aai[0] - aai[2]; | tmr = aar[0] - aar[1]; |
| aar[0] = aar[0] + aar[2]; | tmi = aai[0] - aai[1]; |
| aai[0] = aai[0] + aai[2]; | aar[0] = aar[0] + aar[1]; |
| aar[2] = tmr; | aai[0] = aai[0] + aai[1]; |
| aai[2] = tmi; | aar[1] = tmr; |
| windex = 1; | aai[1] = tmi; |
| //j=1 | index = 2; index2 = 3; |
| C = wr[1]; S = wi[1]; | tmr = aar[2] - aar[3]; |
| index = 1; index2 = 3; | tmi = aai[2] - aai[3]; |
| tmr = aar[1] - aar[3]; | aar[2] = aar[2] + aar[3]; |
| tmi = aai[1] - aai[3]; | aai[2] = aai[2] + aai[3]; |
| aar[1] = aar[1] + aar[3]; | aar[3] = tmr; |
| aai[1] = aai[1] + aai[3]; | aai[3] = tmi; |
| aar[3] = tmr*C – tmi*S; | |

Figure 14. List of assignments after symbolic simulation, removing controlling variables and adjusting array indexes.

In other words, dynamic CP insertion increases the scalability of M-HED for addressing equivalence checking problem. Selecting the right number and location of CPs is a challenge. While choosing too few CPs leads to a blow up of the forward M-HED construction, choosing too many CPs results in lots of re-substitutions for false negative elimination. In our proposed method, in an iterative way, a CP is inserted in a location so that the size of corresponding M-HED becomes as large as possible. This way, two segments are generated.

One segment in which the corresponding M-HED size is equal to the maximum allowed M-HED size and the other one in which the corresponding M-HED size can be even larger than the maximum allowed M-HED size. The first generated segment of the specification (*sasc*) and the implementation (*sprtl*) are compared to check for equivalency. Although in HLS due to resource sharing, the mapping between specification and RTL implementation usually is many to one, it is not necessary to compare each node in the segment of specification with all nodes in the segment of RTL implementation. That is because we make use of a canonical decision diagram, i.e. M-HED, so that two nodes that are functionally equivalent are automatically detected. In each iteration of the *SEC-PIPED* algorithm, the following operations are done.

If *LAASC* is not empty (line 6 of Figure 15), we insert a CP in the assignment list. As a result, two segments are generated. The size of corresponding M-HED of the first segment (*sasc*) is equal to the maximum size allowed by M-HED. A segment in *LAASC* (*sasc*) is chosen by *segmentSelector* function and removed from *LAASC* (lines 7-8). Next, M-HED of all statements in *sasc* is created (*Hsasc* in line 9). In the same way, if *LAPRTL* is not empty (line 13), a CP is inserted in *LAPRTL*. Then a segment in *LAPRTL* (*sprtl*) is chosen by using *segmentSelector* function again and removed from *LAPRTL* (lines 14-15). Next, M-HED representation of *sprtl* (*Hsprtl*) is created (line16).

Please note that the maximum size of a design that can be handled by M-HED is dependent on the structure of design. Typically, M-HED can handle 30000 assignment lines. Based on this information as well as the structure of designs to be verified the proper location of CP is automatically determined. After creating *Hsasc* and *Hsprtl,* they are compared using M-HED and the equivalent output nodes are specified by *eo* (line 20). As mentioned in Section 4-1, the output nodes are those that appear only the left-hand side in a segment. As a result of equivalence checking, the equivalent output nodes as well as those nodes that equivalent outputs are dependent on them (*dni* and *dns*) are removed from *sasc* (lines 24-26) and *sprtl* (lines 29-31). New primary inputs for equivalent nodes in their places are introduced that are used in the next segments of *LAASC* and *LAPRTL*. Note that remaining segments are updated using *UpdateSegment* function in lines 27 and 32 in order to reflect the effect of these new primary inputs.

---

**SEC-PIPED (ASC, PRTL)**

1  LAASC  = SymSim(ASC);
2  LAPRTL= SymSim(PRTL);
3  sasc  = ∅;
4  sprtl = ∅;
5  WHILE (LAASC ≠ ∅ or LAPRTL ≠ ∅)
6     IF (LAASC ≠ ∅)
7        sasc = segmentSelector(LAASC);
8        LAASC = LAASC – sasc;
9        Hsasc = HEDGen(sasc);
10    ELSE
11       Hsasc = HEDGen(sasc);//since it isn't constructed in Line9
12    END IF
13    IF (LAPRTL ≠ ∅)
14       sprtl = segmentSelector(LAPRTL);
15       LAPRTL = LAPRTL – sprtl;
16       Hsprtl = HEDGen(sprtl);
17    ELSE
18       Hsprtl = HEDGen(sprtl);//since it isn't constructed in Line16
19    END IF
20    eo = EquChecking (Hsasc, Hsprtl);  //equivalent output nodes;
21    IF (eo= ∅) //no equivalent nodes are found
22          (LAASC, LAPRTL, sasc, sprtl) =
                  INTERNAL-EQU(LAASC,  LAPRTL, sasc, sprtl);
23    ELSE
24       sasc = sasc – eo;//remove equ. output nodes from sasc
25       dns = {n ∈ sasc: ∃ eq ∈ eo & n has an impact on eq};
26       sasc = sasc – dns;
27       LAASC=UpdateSegment(LAASC, eo, dns, PIs)
28       sprtl = sprtl – eo;//remove equ. output nodes from sprtl
29       dni = {n ∈ sprtl: ∃ eq ∈ eo & n has an impact on eq};
30       sprtl = sprtl – dni;
31       LAPRTL =UpdateSegment(LAPRTL, eo, dni, PIs);
32    END IF
33 END WHILE
34 IF (sasc = ∅ and sprtl = ∅)
35     RETURN "EQUIVALENT Designs";
36 ELSE
37     RETURN "UNEQUIVALENT Designs";
38 END IF

---

Figure 15. Sequential equivalence checking (SEC) of pipelined data path designs.

Another point to be noted here is that removing internal nodes when they have impact on equivalent nodes is not always safe. That is because these nodes may also have impact on other nodes that are checked yet. Replacing them with primary inputs would remove useful correlation among other nodes. To avoid this unsafe operation, during updating remaining segments by *UpdateSegment* function, if internal nodes are used in unprocessed segments, they are described in terms of primary inputs. To do so, we utilize useful embedded feature of M-HED described in Section 3 that help us to do it easily.

It should be noted that the *while* loop of *SEC-PIPED* algorithm is finished when all segments in both *LAASC* and *LAPRTL* become empty. If the last updated *sasc* and *sprtl* are empty, it means that for all output nodes in *sasc*, there is an equivalent node in *sprtl* and vice versa which necessitate the specification and implementation to be equal (lines 34-35). Otherwise they are not equivalent (line 37).

Figure 16 illustrates the main idea behind *SEC-PIPED* algorithm. In this figure, $p_i$, $i_i$, $s_i$ and $np_i$ are primary input, implementation node, specification node, and finally new primary input defined in the place of equivalent nodes. As shown in this figure, the first segment of the implementation (*sprtl*) and specification (*sasc*) are compared using M-HED. The equivalent output nodes of these segments ($i_4$, $i_5$, $s_1$, $s_2$) are detected and segments are updated by introducing new primary inputs in the place of equivalent nodes ($np_1$ and $np_2$ Figure 16(b)). Besides, during equivalence checking nodes that impact on equivalent nodes are removed except nodes that may affect on other nodes to be checked ($i_2$ in Figure 16(a)). These nodes are detected during updating segments and described in polynomial form in terms of primary inputs using the useful property of M-HED ($p_2 p_3$ in Figure 16(b)). Section 3 described how this polynomial can be written. The new segments are compared again and equivalent output nodes ($i_8$ and $s_3$ in Figure 16(b)) are detected.

### 4.3 Finding Internal Equivalent Nodes

Suppose we follow the *SEC-PIPED* algorithm, but in the given segment size, no equivalent nodes exist (lines 21-23 of Figure 15). In this case no nodes can be removed from *sasc* or *sprtl* and therefore M-HED construction is blocked. Since the size of the corresponding M-HED of *sasc* or *sprtl* is equal to the maximum allowed M-HED size, in the next iteration no statement can be added to the selected segment and therefore this segment is selected again and again and the algorithm falls into an endless loop. In fact, because of several redundancies or optimizations added during pipelined RTL generation, the size of assignment lists of specification (*LAASC*) and implementation *(LAPRTL)* would be different. Hence, when we choose one pair segment for comparison (*sasc* and *sprtl*) equivalent output nodes may not exist. The question that rises in the mind of reader is whether or not our methodology can handle such a case.

In order to avoid blocking forward M-HED construction, *INTERNAL-EQU* algorithm shown in Figure 17 is proposed. The basic idea is to look for non-output (internal) equivalent nodes. To do so, first we compare output nodes of *Htsprtl* and a modified version of *Htsasc* while output nodes of *Htsprtl* and the original *Htsasc* are not equivalent. To obtain the modified *Htsasc*, *sasc* and *sprtl* are reserved in temporary locations (*tsasc* and *tsprtl* in lines 1-2 of Figure 17). We omit output statements of *sasc*, and append these statements to *LAASC* (lines 9-11 of Figure 17). After such deletions, some previously intermediate nodes appear as new output nodes. Afterwards, the M-HED of the modified *tsasc* is constructed (line 12). If all statements are removed and we are not able to obtain any new equivalent nodes, the process is repeated for *tsprtl* (lines 17-24). This is because output nodes of *sprtl* may be equivalent to internal nodes of *sasc* or vice versa. To perform such an operation, first, *sasc* is retrieved and the corresponding M-HED is reconstructed (lines 15-16). Then all processes of the first while loop is repeated for *sprtl* to obtain internal nodes which are equivalent to some output nodes of *sasc*. In this algorithm, when an equivalent pair of nodes is found, *sasc*, *sprtl*, *LAASC*, *LAPRTL* are updated and returned as results (lines 7 and 19). This way, we are able to detect equivalent nodes based on an iterative deletion approach.

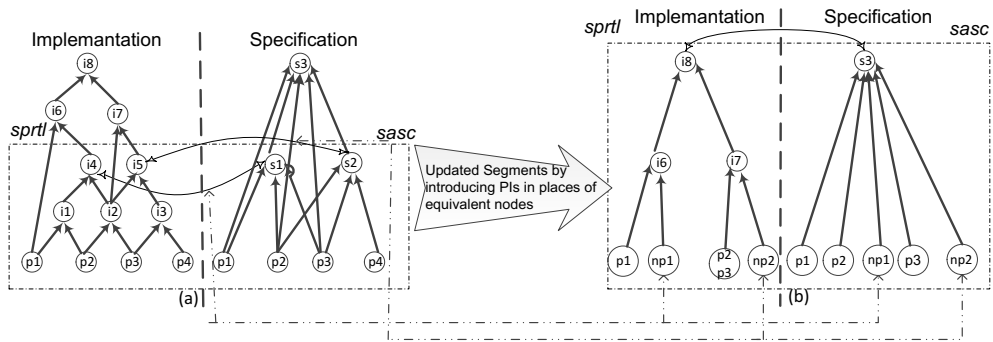

Figure 16. Illustrations of the main part of SEC-PIPED function (a) when the first segments (*sasc* and *sprtl*) are compared (b) when the second segments (*sasc* and *sprtl)* are compared.

| **INTERNAL-EQU(LAASC, LAPRTL, sasc, sprtl)** |
|---|
| 1        *tsasc =sasc;* |
| 2        *tsprtl =sprtl;* |
| 3        *Htsasc = HEDGen(tsasc);* |
| 4        *Htsprtl = HEDGen(tsprtl);* |
| 5    **WHILE** *(tsasc≠ ∅)* |
| 6      **IF** *EquChecking (Htsasc,Htsprtl) ≠ ∅* |
| 7              **RETURN** *LAASC, LAPRTL, tsasc, tsprtl;* |
| 8      **ELSE** |
| 9        *ontsasc = outputs statements in tsasc;* |
| 10      *tsasc   = tsasc - ontsasc;* |
| 11      *LASASC = LAASC ∪ ontsasc;* |
| 12      *Htsasc = HEDGen(tsasc);* |
| 13    **END IF** |
| 14  **END WHILE** |
| 15  *tsasc = sasc* |
| 16  *Htsasc = HEDGen(tsasc);* |
| 17  **WHILE** *(tsprtl≠ ∅)* |
| 18    **IF** *EquChecking (Htsasc,Htsprtl) ≠ ∅* |
| 19            **RETURN** *LAASC, LAPRTL, tsasc, tsprtl;* |
| 20    **ELSE** |
| 21          *ontsprtl = outputs statements in tsprtl;* |
| 22      *tsprtl   = tsprtl- ontsprtl;* |
| 23      *Htsprtl  = HEDGen(tsprtl);* |
| 24      *LAPRTL=LAPRTL ∪ ontsprtl;* |
| 24    **END IF** |
| 25  **END WHILE** |
| 26  **RETURN** *"UNEQUIVALENT Designs"* |

Figure 17. Procedure of finding internal equivalent nodes.

Figure 18 demonstrates how to use this algorithm. Suppose that the maximum size of M-HED only allows us to construct eleven nodes of implementation and six nodes of specification. As shown in Figure 18(a), because $s_1$ and $s_2$ are equivalent to $i_6$ and $i_7$, and also $i_4$ and $i_5$ are considered as internal nodes (output nodes are $i_6$ and $i_7$), no output nodes are equivalent. Hence, by using *INTERNAL-EQU* algorithm, output nodes of *sprtl* ($i_6$ and $i_7$) are removed and internal nodes ($i_4$ and $i_5$) are introduced as new outputs. This way, equivalent output nodes are appeared and constructing M-HED can be resumed similar to Figure 16(a) as seen in Figure 18(b).

## 4.4 Example

In this subsection we illustrate our methodology with an example. As mentioned before, Figure 5 and Figure 6 show a nested loop in C code as an *ASC* block and related synthesized hardware. As you can see in Figure 5, in the *ASC* the results of two multiplications are added and stored in temporary variables (*tempf* and *temps*). These temporary variables are multiplied and stored in a *res* array. The RTL code synthesized by CatapultC is not pipelined and therefore finding cut-loop is possible.

However, Figure 6 indicates a sequence of cycles when a RTL code with a loop pipelined is synthesized. As mentioned in Section 2, pipelining makes the equivalence checking hard. Figure 19 demonstrates the three steps of equivalence checking using our methodology. In this figure, output nodes in each segment of implementation and specification are colored as blue and green respectively. Suppose that the first segment of the specification (*segment_{1s}*) includes *tempf0* and *temps0* as output nodes and the first segment of the implementation (*segment_{1i}*) includes *mul2*, *mul3*, and *add0* as output nodes. After creating M-HEDs for all assignments in these segments, *add0* ∈ *segment_{1i}* and *tempf0* ∈ *segment_{1s}* are detected as equivalent nodes. However, because *add1* ∉ *segment_{1i}*, neither *add1* nor any other node is detected as an equivalent node to *temps0*. At this point, a new primary input is defined instead of equivalent nodes (NEW PI in Figure 19(b)) and {*mul0*, *mul1*, *add0*} ∈ *segment_{1i}* and {*tempf0*} ∈ *segment_{1s}* are removed.

Since {*mul1*, *mul2*} ∈ *segment_{1i}* have no impact on *add0* and {*temps0*} ∈ *segment_{1s}* has no equivalent node in *segment_{1i}*, they are not removed from *segment_{1i}* and *segment_{1s}*. In the next phase, as shown in Figure 19(b), new segments *segment_{2s}* and *segment_{2i}* are taken into account so that {*NewPI, temps0, res00*} ∈ *segment_{2s}* and { *NewPI, mul2, mul3, mul4, mul5, add1*} ∈ *segment_{2i}*. During equivalence checking, no output node is matched and therefore *INTERNAL-EQU* algorithm needs to be called. In this algorithm, first of all, output *res00* is removed from *segment_{2s}* which makes *temps0* a primary output of *segment_{2s}* (Figure 19(c)). Then, M-HEDs of updated segments are constructed again and *SEC-PIPED* is resumed. At this point, the equivalent nodes *temps0* and *add1* are detected. Next, they are removed and new primary inputs are defined instead of them. This procedure is continued so that all nodes can be removed from *sasc* and *sprtl*. If all segments have been processed, *LAASC* and *LAPRTL*, and also the final *sasc* and *sprtl* have become empty, the algorithm returns *EQUIVALENT Designs*. As it can be observed, our proposed solution can be applied to RTL designs with pipelined nested loops and complex structure.

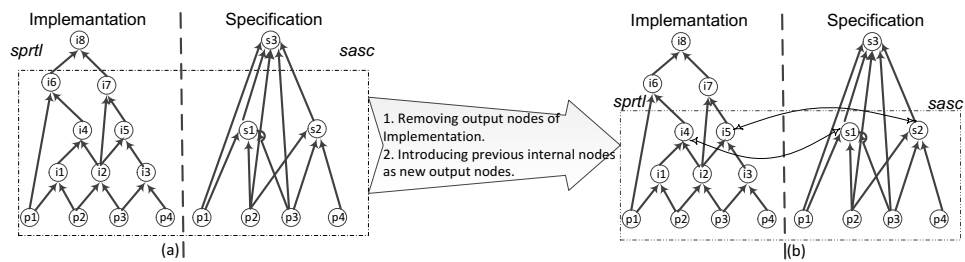

Figure 18. Illustration of the main part of INTERNAL-EQU function. (a) before finding internal-equivalent nodes when the first segments (*sasc* and *sprtl*) are compared. (b) after finding internal-equivalent nodes when the first segments (*sasc* and *sprtl*) are compared.

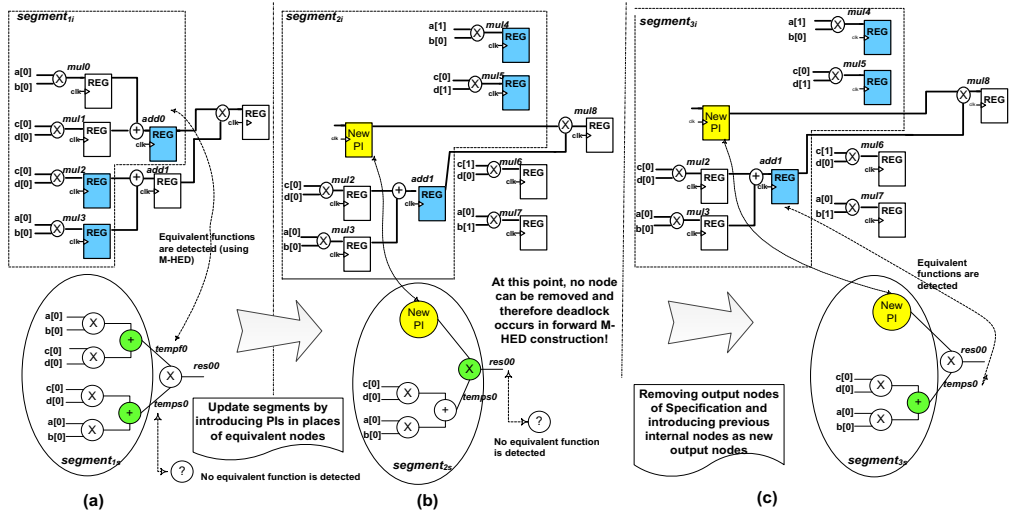
Figure 19. Demonstration of steps of our proposed equivalence checking methodology

## 5. LIMITATIONS OF PROPOSED METHODOLOGY

False negatives are failing properties that unintentionally raise a wrong flag as a sign of unequal design while there is no bug in the design. Verification engineer must eliminate false negatives to make sure only real bugs lead to failing properties [19]. It is well known that using each type of cut-points in each kind of decision diagrams and even equivalence checking approaches can potentially lead to false negatives. In the proposed methodology we try to avoid false negatives in different ways.

First, we move nonequivalent nodes in a given segment $i$ into another segment $j$ and repeat the checking process until nodes become equivalent or all segments are covered (i.e. at the end of statements). In fact, our methodology doesn't decide about the results of equivalence checking immediately after finding nonequivalent nodes in a pair of checked segments. Because two nodes in a specific *sasc* and *sprtl* (e.g. $sasc_i$ and $sprtl_k$) may not be equivalent, while by moving a node from $sasc_i$ into $sasc_j$, we can find equivalent nodes (like *temps0* in Figure19). The scalability of the proposed methodology can be kept by removing equivalent nodes and nodes that effect on them. These eliminations reduce the size of equivalence checking significantly.

Second, whenever internal nodes those effect on equivalent nodes as well as other nodes in the unprocessed segments are removed, new inputs are introduced instead of them. These inputs are described in terms of primary inputs of design in polynomial form using M-HED that provides facilities to avoid false negative results as much as possible (like $i_s$ in the Figure16(a) ). Note that false positives do not occur on M-HED due to its canonical representation.

## 6. EXPERIMENTAL RESULTS

Our methodology for equivalence checking has been implemented in C++ and carried out on an Intel 2.8 GHz Corei7 with 8 GB main memory running Linux with Qt creator as an IDE. In order to demonstrate the effectiveness of the proposed equivalence checking technique, we apply our technique to several designs common in Digital Signal Processing (DSP) and multimedia applications. The benchmarks include *ColorConversion* as an algorithm enabling different conversion standards to be supported with the same hardware, *Sobel* as a convolution algorithm which is the core of many image processing algorithms, Finite Impulse Response with two sizes *(FIR4, FIR32)* as the most common digital filter, Discrete Cosine Transform (*DCT*16, *DCT*32) and Fast Fourier Transform (*FFT32, FFT64, FFT256, FFT512*). These designs come from a variety of problem domains such as mathematics, digital signal processing and multimedia. Furthermore, we employ CatapultC as an HLS tool to automatically generate RTL codes from C code of these benchmarks[1]. Each circuit has been piped in a certain frequency. In addition we use Minisat [35] as a SAT-solver and Z3 [36] which is generally considered the fastest SMT solver. The time out *(TO)* is set to 1000 seconds.

Table I shows the experimental results with and without using our methodology. The first column (*Benchmark*) is the benchmark name. The second column (*#LRTL*) shows the number of RTL lines of each benchmark generated by CatapultC. The Third column *(#LS)* shows the number of lines obtained after symbolic simulation. *Loop information* column indicates the loop information such as the number of single, 2-nested and 3-nested loops in the design. The major column *without our methodology* shows the results when primary outputs are directly expressed in terms of primary inputs and then represented by M-HED. The last major column, *with our methodology*, shows the results in terms of memory usage (*MemoryUsage*) and required processing time (*CpuTime*) after applying the proposed method.

Table I. Experimental results of equivalence checking without/with our methodology.

| Benchmark | # LRTL | #LS | Loop information | | | without our methodology | | with our methodology | |
|---|---|---|---|---|---|---|---|---|---|
| | | | Singles | 2-nested | 3-nested | MemoryUsage (MB) | CpuTime (s) | MemUsage (MB) | CpuTime (s) |
| ColorConversion | 141 | 712 | 0 | 1 | 0 | 1 | 0.1 | 1 | 0.1 |
| FIR4 | 170 | 848 | 2 | 0 | 0 | 1.9 | 0.2 | 1.9 | 0.2 |
| FIR32 | 243 | 2212 | 2 | 0 | 0 | 5.2 | 6.1 | 2.6 | 1.3 |
| DCT16 | 854 | 3996 | 0 | 0 | 2 | MO | NA | 5.2 | 6.1 |
| DCT32 | 1258 | 6352 | 0 | 0 | 2 | MO | NA | 11.7 | 9.2 |
| Sobel | 2268 | 9765 | 0 | 3 | 0 | MO | NA | 12.1 | 12.7 |
| FFT32 | 2742 | 10156 | 0 | 0 | 1 | MO | NA | 15.6 | 14.2 |
| FFT64 | 2986 | 14258 | 0 | 0 | 1 | MO | NA | 38.5 | 31.1 |
| FFT256 | 3378 | 32568 | 0 | 0 | 1 | MO | NA | 89.5 | 82.7 |
| FFT512 | 3976 | 88682 | 0 | 0 | 1 | MO | NA | 121.6 | 219.2 |

(MO : Out of 8GB memory; NA : Not applicable, due to the memory out, time is not reported; CPU time is given in seconds)

As the results show, in some cases (*ColorConversion, FIR4, and FIR32*) *without our methodology* and *with our methodology* can handle the problem of equivalence checking, but in other cases our methodlogy can handle it efficiently, while *without our methodology*, we have faced with memory out (*MO*) problem. These results convince us that equivalence checking without our methodology is prohibitively expensive and even impossible for large designs. Furthermore, as stated in Section 2, directly using of several optimizations for equivalence checking such as cut-loop and cut-plane techniques is inapplicable when the loops of thr design is pipelined. In this situation, our methodology can solve the problem of equivalence checking efficiently. In fact, M-HED can represent arithmetic operations at word level representation and there is no need to encode them to bit-level operations. Besides, it can handle bit-level operations as well as world level. Indeed M-HED is a strong and scalable decision diagram for representing and verification of datapath circuits [27, 30]

In another experiment, we have tried to solve the equivalence checking by using a SAT-solver. The results reported in Table II show that using SAT-solvers in datapath circuits especially with many arithmetic components is inefficient. As it can be seen, using SAT increases verification run time significantly even for small circuits. Obviously, when using our methodology, the run time for equivalence checking is reduced by 111.9× on average, i.e., two orders of magnitude of average speedup. As opposed to low level methods such as Boolean SAT based techniques, the results indicate that our method not only uses an efficient canonical form to represent symbolic expressions but also is scalable even on large circuits.

In order to complete the set of results, we compared our results against using Z3 as an SMT-solver in Table III. SMT-solvers try to handle the weakness of SAT-solvers on arithmetic designs by combining SAT-solvers with different mathematical theories. In these engines, the input design is first simplified by using different theories such as linear arithmetic, theory of arrays, and bit-vectors approaches.

Table II. Improvements in comparison with SAT-based method

| Benchmark | Using SAT | | Improvements by using our methodology | |
|---|---|---|---|---|
| | MemoryUsage (MB) | CpuTime (sec) | MemoryUsage (MB) | CpuTime (sec) |
| ColorConversion | 9.2 | 41.1 | 9.2× | 411× |
| FIR4 | 10.1 | 54.6 | 5.3× | 273× |
| FIR32 | 28.2 | 121 | 10.9× | 93.1× |
| DCT16 | 32.2 | 301.2 | 6.2× | 49.4× |
| DCT32 | 47.2 | 518.6 | 4.1× | 56.4× |
| Sobel | 78.2 | 872.7 | 6.5× | 68.7× |
| FFT32 | 87.2 | 923.5 | 5.6× | 65.1× |
| FFT64 | NA | TO | NA | 48.2× |
| FFT256 | NA | TO | NA | 36.3× |
| FFT512 | NA | TO | NA | 18.3× |
| Average Improvement by using our methodology | | | NA | 111.9× |

TO: Out of 1000 sec; NA : Not applicable, due to the timeout, the memory usage is not reported

Hence, the remaining instance is smaller and easier to solve. Although the most useful theories in equivalence checking are related to bit-vector and array, in small cases, using these theories for abstraction and simplification versus solving the problem by M-HED is a time-consuming task and therefore the run time increases. For *DCT*16 and *DCT*32 benchmarks, the results obtained by SMT-solver are better than those of M-HED. That is because, in these benchmarks, the synthesized designs have many bit-level descriptions. Although M-HED has features to handle bit-level operations efficiently, SMT-solvers are strong and powerful tool for verification when word-level and many bit-level operations are mixed in a single design. In comparison with M-HED, they are not good enough to handle datapath designs which are described mostly in word level. In addition, without using our methodology, word level engines cannot handle the equivalence checking problem of datapath pipelined loop design efficiently.

Table III. Improvements in comparison with SMT-based method.

| Benchmark | Using SMT | | Improvements by using our methodology | |
|---|---|---|---|---|
| | MemoryUsage (MB) | CpuTime (sec) | MemoryUsage (MB) | CpuTime (sec) |
| ColorConversion | 1.5 | 1.2 | 1.5× | 12× |
| FIR4 | 2.5 | 1.7 | 1.3× | 8.5× |
| FIR32 | 2.8 | 3.5 | 1.1× | 2.7× |
| DCT16 | 3.9 | 4.4 | 0.8× | 0.7× |
| DCT32 | 7.7 | 6.5 | 0.7× | 0.7× |
| Sobel | 19.8 | 15.1 | 1.6× | 1.2× |
| FFT32 | 44.3 | 21.8 | 2.8× | 1.5× |
| FFT64 | 101.9 | 47.2 | 2.6 | 1.5× |
| FFT256 | MO | NA | 89.4× | NA |
| FFT512 | MO | NA | 65.8× | NA |
| Average Improvement by using our methodology | | | 16.7× | NA |

TO: MO : Out of 8GB memory; NA : Not applicable, due to the memoryout, the time is not reported

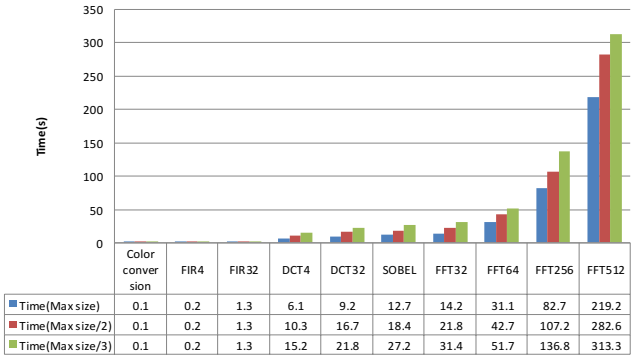

| | Color conversion | FIR4 | FIR32 | DCT4 | DCT32 | SOBEL | FFT32 | FFT64 | FFT256 | FFT512 |
|---|---|---|---|---|---|---|---|---|---|---|
| Time(Max size) | 0.1 | 0.2 | 1.3 | 6.1 | 9.2 | 12.7 | 14.2 | 31.1 | 82.7 | 219.2 |
| Time(Max size/2) | 0.1 | 0.2 | 1.3 | 10.3 | 16.7 | 18.4 | 21.8 | 42.7 | 107.2 | 282.6 |
| Time(Max size/3) | 0.1 | 0.2 | 1.3 | 15.2 | 21.8 | 27.2 | 31.4 | 51.7 | 136.8 | 313.3 |

Figure 20. The effect of segmentation size on the run time.

In the last experiment, the effect of segmentation size and selecting the right number and location of CPs on the run time is investigated. Figure 20 reports the results for three cases: maximum allowed M-HED size (Max Size), Max Size/2 and Max Size/3. As can be seen in this figure, the running time has remained fixed in the first three cases. This is because the whole design is located in one segment, even in the case of Max Size/3. The results show that by reducing the segmentation size the processing time increases. That is because by reducing the segmentation size the number of non-equivalent output nodes in each segment increases and therefore *INTERNAL-EQU* function must be used to avoid blocking forward M-HED construction and therefore the processing time increases.

## 7. CONCLUSION

In this paper, we have introduced a formal equivalence checking methodology for behavioral synthesized pipelined designs with nested loops based on a canonical decision diagram called M-HED that supports modular polynomial computations. For increasing the scalability of our methodology, we employ dynamic cut-points which enable us to effectively perform sequential equivalence checking. To the best of our knowledge, this is the first work that formally checks the equivalence of pipelined nested loops by using high level decision diagrams. The experimental results demonstrate that our proposed methodology can support designs with arbitrary structures and large data path. Average improvements in terms of the memory usage and run time in comparison with SMT- and SAT-based equivalence checking are 16.7× and 111.9× respectively.